\documentclass{article}

\usepackage{arxiv}

\usepackage[toc,page]{appendix}

\usepackage{graphicx}
\usepackage{dcolumn}
\usepackage{bm}

\usepackage[mathlines]{lineno}

\usepackage[utf8]{inputenc}
\usepackage{gensymb} 
\usepackage{booktabs} 
\usepackage{subcaption} 
\usepackage{color} 

\usepackage{amsmath}

\usepackage[utf8]{inputenc}
\usepackage[english]{babel}
\usepackage{graphicx}
\usepackage{gensymb} 
\usepackage{booktabs} 

\usepackage{lipsum}

\usepackage[normalem]{ulem} 
\usepackage{bm}
\usepackage{bbm}
\usepackage{amssymb}

\usepackage{url}

\usepackage{tikz}
\usetikzlibrary{shapes.geometric}

\usepackage{algcompatible}
\usepackage{newfloat}
\DeclareFloatingEnvironment[
    fileext=loa,
    listname=List of Algorithms,
    name=ALGORITHM,
    placement=tbhp,
]{algorithm}

\title{Optimization of patient-specific range modulators for conformal FLASH proton therapy}

\usepackage{authblk}

\author[1]{Sylvain Deffet}
\author[1,2]{Kevin Souris}
\author[1,3,4]{Edmond Sterpin}

\affil[1]{ 
Université catholique de Louvain, Institut de Recherche Expérimentale et Clinique (IREC), Center of Molecular Imaging, Radiotherapy and Oncology (MIRO), Louvain-La-Neuve, 1348, Belgium
}%
\affil[2]{Ion Beam Applications SA, Louvain-La-Neuve 1348, Belgium}
\affil[3]{KU Leuven, Department of Oncology, Laboratory of Experimental Radiotherapy, Leuven, 3000, Belgium}
\affil[4]{Particle Therapy Interuniversity Center Leuven - PARTICLE, Leuven, 3000, Belgium}

\date{\today}

\renewcommand{\shorttitle}{FLASH dose rate}

\begin{document}
\maketitle

\begin{abstract}
\textbf{Background:} In proton therapy, current pencil beam scanning (PBS) systems cannot deliver intensity modulated proton therapy (IMPT) treatment with a FLASH dose rate. A promising approach to enable FLASH conformal proton therapy is to passively degrade a single energy layer using a patient-specific range modulator. This range modulator can be seen as a combination of a ridge filter and a range shifter to achieve both uniformity and conformality. Several studies have already proved the feasibility of this approach. However, in those published works, the optimization of the range modulators is more akin to dose mimicking as it is not performed with respect to the constraints used to design the original IMPT plan. In addition, a complex simulation pipeline with an external dose engine is required to deal with the parameterized geometries of the range modulators.\\
\textbf{Purpose:} We propose an innovative method to directly optimize the geometrical characteristics of the range modulator and the treatment plan with respect to user defined constraints, similarly to state-of-the-art IMPT inverse planning.\\
\textbf{Methods:} The kind of range modulators proposed in this study is a voxelized object which can be placed in the CT for dose computation, which simplifies the simulation pipeline. Both the geometrical characteristics of the range modulator and the weights of the PBS spots were directly optimized with respect to constraints on the dose using a first-order method. A modified Monte Carlo dose engine was used to provide an estimate of the gradient of the relaxed constraints with respect to the elevation values of the range modulator.\\
\textbf{Results:} Assessed on a head and neck case, dose conformity logically appeared to be significantly degraded compared to IMPT. We then demonstrated that this degradation came mainly from the use of a large range shifter and therefore from physical limitations inherent in the passive degradation of beam energy. The geometry of the range modulator, on the other hand, was shown to be very close to being optimal. PBS dose rates were computed and discussed with respect to FLASH objectives.\\
\textbf{Conclusions:} The voxelized range modulators optimized with the proposed method were proven to be optimal on a head and neck case characterized by two rather large volumes, with irregular contours and variable depths. The optimized geometry differed from conventional ridge filters as it was arbitrarily set by the optimizer. This kind of range modulators can be directly added in the CT for dose computation and is well suited for 3D printing.
\end{abstract}


\section{Introduction}

Recent studies have shown that the use of very high dose rates would improve the protection of healthy tissues without compromising tumor control\cite{Favaudon2014, BUONANNO2019, MontayGruel2019}. However, the preclinical results are sometimes contradictory\cite{Venkatesulu2019, BEYREUTHER2019} and the underlying mechanism is not well understood yet\cite{Pratx2019, Spitz2019, LABARBE2020, JIN2020}. Preliminary findings suggest that a dose rate of at least 40 Gy/s is a critical prerequisite for the manifestation of the FLASH effect\cite{Favaudon2014}. Nevertheless, recent research indicates that this threshold may not be universal, and the degree of the effect may depend on the dose rate\cite{MONTAYGRUEL2017, Rothwell2022}. It also seems that there could be a minimum dose threshold\cite{Rothwell2022}.

State-of-the-art proton therapy relies on pencil beam scanning (PBS) to achieve excellent dose conformity. However, current PBS systems cannot deliver an intensity modulated proton therapy (IMPT) treatment with a FLASH dose rate. On the one hand, changing the beam energy from one layer to the other takes about 1 second. Consequently, the time to deliver a field can be around 20 seconds. On the other hand, several fields are generally required and the treatment can be fractionated over several days.

In the light of the above, current PBS systems must be adapted to enable FLASH treatments. The combined use of a single energy layer and a patient-specific range modulator would decrease the irradiation time and thus increase the dose rate while maintaining conformity, with minimal system modifications\cite{Rothwell2022}. The 3D range modulator typically consists of a collection of pyramids of different heights which transforms the pristine Bragg peaks into spread-out ones. The base of the range modulator is shaped to act as a range compensator to match the distal contour of the tumor. It has been shown that this set-up could easily achieve dose rate of at least 40 Gy/s\cite{Zhang2022, Liu2022}.

To plan a conformal PBS FLASH treatment it is therefore necessary to optimize a patient-specific range modulator and the weights of the PBS spots. Several methods have already been proposed\cite{Simeonov2022, Liu2022, Zhang2022}. Although different, they are based on two common principles:
\begin{enumerate}
    \item A Monte Carlo dose engine is used to create a dictionary of dose distributions from which the sizes of the slabs of the pyramids constituting the range modulator are computed. To this end, a reference IMPT plan is first optimized and the energy levels are converted into slabs of range shifter.
    \item The optimized range modulator is placed between the nozzle exit and the CT to calculate the dose influence matrix required for spot weight optimization.
\end{enumerate}

These studies have proved the feasability of using a single energy layer with a range modulator to deliver a dose to a complex target volume. However, one can regret that the optimization of the range modulators is akin to dose mimicking as it is not performed with respect to the constraints of the original IMPT plan. In addition, this optimizations require a complex simulation pipeline with an external dose engine.

We present a new approach which is different in two aspects.

First, the modulator geometry and spot weights are optimized directly with respect to dose constraints similar to state-of-the-art IMPT inverse planning. We assume that the geometry of the range modulator and the weights of the PBS spots are intrinsically linked and benefit from not being optimized separately. In a way, this is similar to IMPT optimization where it is not appropriate to select the final entries of the dose influence matrix before optimizing the spot weights. The best solution can only be obtained when considering all possible combinations of range modulators and spot weights.

Secondly, we do not use any priors on the geometry of the range modulator. In particular, we do not impose that it consists of pyramids. On the contrary, our range modulator has a pixelated geometry. It can be represented by an elevation map in raster form. Thus, the optimizer generates an optimal arbitrary geometry consisting eg. of truncated and non-symmetrical pyramids to better conform to the complex shape of the target volume. In addition, with such a voxelized geometry the range modulator can directly be added in the CT to perform dose calculations without having to resort to a tool supporting parameterized geometries.

The approache presented in this paper was applied to a head and neck case and compared to state-of-the-art IMPT.

\section{Methods}
\subsection{Overview of the treatment plan optimization process}
\begin{figure}[h]
    \centering
    \includegraphics[width=0.45\textwidth]{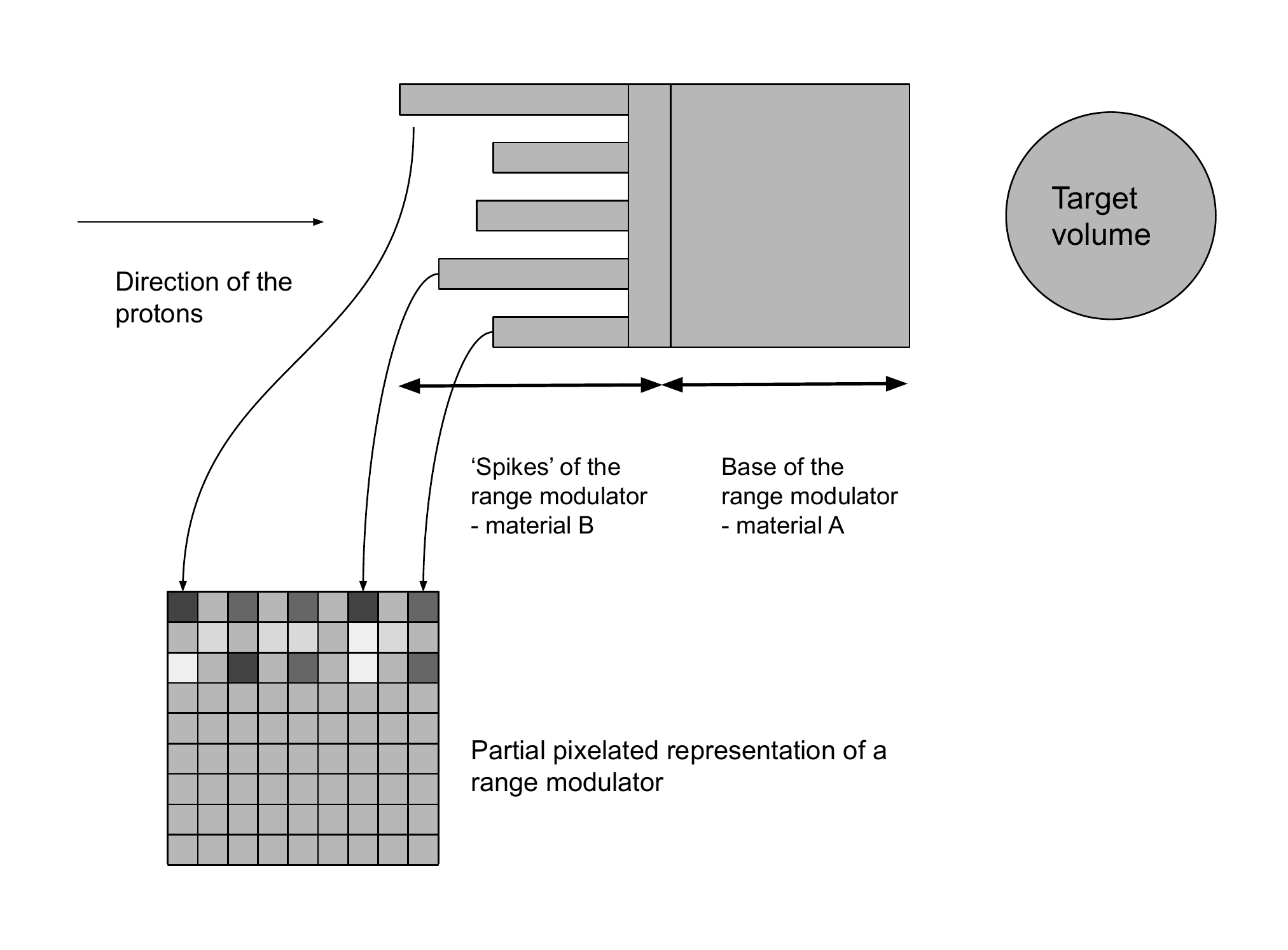}
    \caption{A 3D range modulator can be represented as a 2D elevation map. It is split into 'towers' and a range shifter. A minimum thickness of plain material is left attached to the 'towers' accordingly to 3D printing specification.}
    \label{fig:schem_CEF}
\end{figure}

In classical inverse planning, the parameters subject to optimization are the spot weights $\bm{w}$. For conformal FLASH, we must also optimize the individual heights, named $\bm{h}$, of the range modulator which is coded as an elevation map as shown on Fig.~\ref{fig:schem_CEF}.

Following a classical approach, the objectives penalize deviations of the dose from a reference value in a region of interest (ROI)\cite{Oelfke2001}. For example, to enforce a minimum dose named $D_{min}$, the following objective term may be minimized:
\begin{equation}
    f_{min}(\bm{w}, \bm{h}) = \frac{1}{N_r} \sum_{i=1}^{N_r} max_{\bm{w}, \bm{h}} \left( 0, D_{min} - D_i(\bm{w}, \bm{h}) \right)^2
    \label{eq:fmin}
\end{equation}
where $N_r = |ROI|$.

A similar objective may be used to enforce a maximum dose named $D_{max}$:
\begin{equation}
    f_{max}(\bm{w}, \bm{h}) = \frac{1}{N_r} \sum_{i=1}^{N_r} max_{\bm{w}, \bm{h}} \left( 0, D_i(\bm{w}, \bm{h}) - D_{max} \right)^2
    \label{eq:fmax}
\end{equation}
%

Several objectives may be defined. Hence, the general treatment plan optimization problem takes the following form:
\begin{eqnarray}
min && \sum_i f_i(\bm{w}, \bm{h})\label{eq:opti_problem}\\
 & s.t. & \bm{w} \geq 0 \nonumber\\
 & & \bm{h} \geq 0\nonumber
\end{eqnarray}
where $f_i(\bm{w}, \bm{h})$ are either of $f_{min}(\bm{w}, \bm{h})$ or $f_{max}(\bm{w}, \bm{h})$ -type.

A common approach for the minimization of Eq.~\ref{eq:opti_problem} with respect to $\bm{w}$ involves a dose influence matrix\cite{Oelfke2001}. Each column of this matrix is the dose contribution of one spot of the treatment plan. Hence, the total dose is computed as a weighted sum of the entries of the dose influence matrix. An analytical formulation of the derivative of the objective terms~\ref{eq:fmin} and~\ref{eq:fmax} can easily be calculated and fed to a first-order optimizer.

On the contrary, the minimization of Eq.~\ref{eq:opti_problem} with respect to $\bm{h}$ cannot be computed using a dose influence matrix. Instead, we propose to compute the derivative of the objetive terms by finite difference as described in details in section~\ref{sec:rm_opti}

\begin{figure}
    \centering
    \begin{tikzpicture}
        \matrix [column sep=7mm, row sep=5mm] {
          \node (initialization) [draw, shape=rectangle, text width=5cm] {Spot grid, RM and weights initialization}; \\
          \node (weightOpti) [draw, shape=rectangle, text width=5cm] {Optimization of the spot weigths $\bm{w}$}; \\
          \node (fluence) [draw, shape=rectangle, text width=5cm] {Estimation of the fluence at RM entrance}; \\
          \node (gradient) [draw, shape=rectangle, text width=5cm] {Estimation of the gradient of the objective function wrt. $\bm{h}$}; \\
          \node (updateH) [draw, shape=rectangle, text width=5cm] {Computation of the next value $\bm{h}$ that decreases the objective function (OF)}; \\
          \node (updateCEM) [draw, shape=rectangle, text width=5cm] {Update RM in CT}; \\
          \node (coninueCEMOpti) [draw, diamond, aspect=2, text width=5cm] {OF can be further decreased wrt. $\bm{h}$ with current $\bm{w}$?};\\
          \node (coninueWOpti) [draw, diamond, aspect=2, text width=5cm] {OF can be further decreased wrt. $\bm{w}$ with current $\bm{h}$?};\\
          \node (done) [draw, diamond] {};\\
        };
        \draw[-latex] (initialization) edge (weightOpti);
        \draw[-latex] (weightOpti) edge (fluence);
        \draw[-latex] (fluence) edge (gradient);
        \draw[-latex] (gradient) edge (updateH);
        \draw[-latex] (updateH) edge (updateCEM);
        \draw[-latex] (updateCEM) edge (coninueCEMOpti);
        \draw[-latex, bend left] (coninueCEMOpti.west) |- node[auto] {yes} (gradient.west);
        \draw[-latex] (coninueCEMOpti) edge node[auto] {no}  (coninueWOpti);
        \draw[-latex, bend right] (coninueWOpti.east) |- node[auto] {yes} (weightOpti.east);
        \draw[-latex] (coninueWOpti) edge node[auto] {no}  (done);
        \end{tikzpicture}
    \caption{Schematics of the iterative optimization of the individual heights of the range modulator (RM) named $\bm{h}$ and the PBS spot weights named $\bm{w}$.}
    \label{fig:optimization_scheme}
\end{figure}
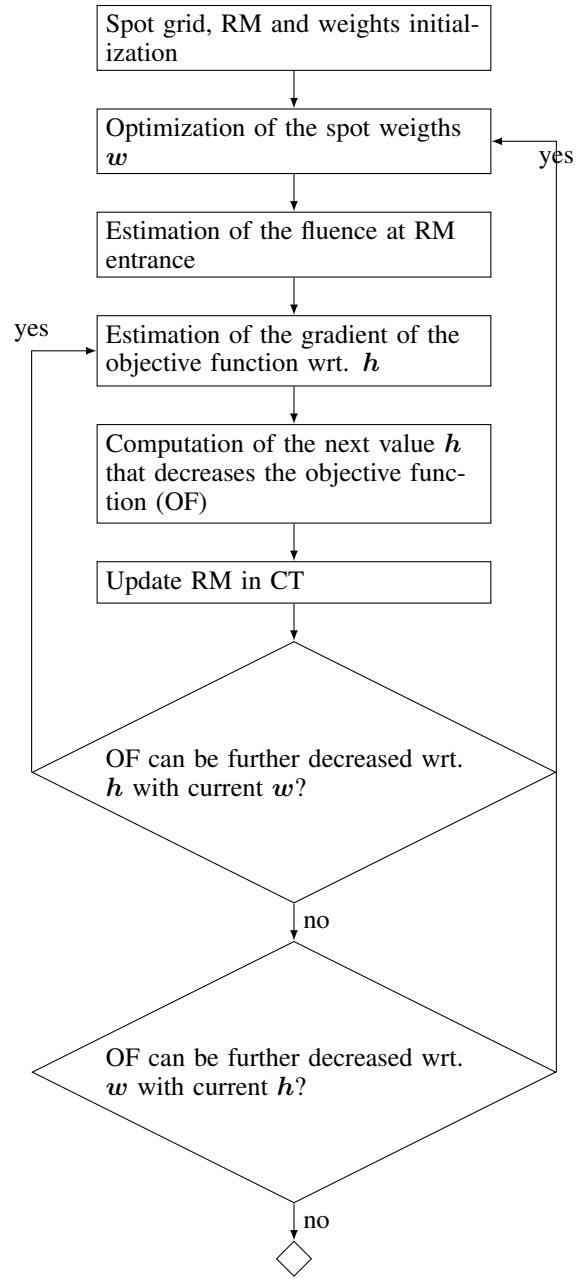

The optimization over $\bm{w}$ and $\bm{h}$ is done iteratively as depicted on Fig.~\ref{fig:optimization_scheme}. Typical layouts are square or hexagonal grids. The optimization of $\bm{h}$ is nested in the optimization of $\bm{w}$. In other words, at each iteration of the optimization of $\bm{h}$, the spots weights $\bm{w}$ are fully re-optimized based on the current value of $\bm{h}$. This choice is a trade-off between the computation times of the two kinds of gradients. The evaluation of the gradient with respect to $\bm{h}$ requires a full computation by finite difference for each update of $\bm{h}$. Conversely, recomputing the gradient with respect to to $\bm{w}$ for new values of $\bm{w}$ is done based on the precalculated dose influence matrix which is must faster.

\subsection{Range modulator optimization\label{sec:rm_opti}}
For each iteration of the range modulator optimization, $\bm{w}$ is fixed and $\bm{h}$ remains the only variable to optimize on.

The gradient with respect to $\bm{h}$ of $f_{min}(\bm{w}, \bm{h})$ is:
\begin{equation}
    \nabla f_{min} = \frac{2}{N_r} \sum_{i=1}^{N_r} max_{\bm{h}} \left( 0, D_{min} - D_i(\bm{w}, \bm{h}) \right) \frac{\delta D_i(\bm{w}, \bm{h})}{\delta\bm{h}}
    \label{eq:grad_fmin}
\end{equation}
A similar gradient may be derived for $f_{max}$.

In Eq.~\ref{eq:grad_fmin}, the computation of $\frac{\delta D_i(\bm{w}, \bm{h})}{\delta\bm{h}}$ is a non trivial operation as the dose $D_i(\bm{w}, \bm{h})$ is usually computed with a Monte Carlo dose engine which does not provide any information of the variation of the particle range with respect to a small variation $\delta E$ of its initial energy. Computing the gradient by finite difference by running a simulation for every element of the vector $\delta \bm{h}$ may be intractable in terms of computation time but also in terms of memory because $|\bm{h}|$ simulations would have to be performed and stored. Therefore, we have adapted our Monte Carlo dose engine to perform and store computation not per beamlet but per pixel of an initial fluence map.

The derivative $\frac{\delta D_i(\bm{w}, \bm{h})}{\delta\bm{h}}$ is not directly estimated. Instead, we compute $\frac{\delta D_i(\bm{w}, \bm{h})}{\delta \bm{E}}$ where $\delta \bm{E}$ is the decrease in energy of the particles at the entrance of the range modulator caused by the increase in thickness $\delta\bm{h}$. It turns out in conformal FLASH that both derivatives are proportional since a single energy layer is used. Specifically, the relationship between the change in energy and in range is given by $\delta E = \delta h \times SP_{RM}(E)$, where $SP_{RM}(E)$ is the stopping power of the material used for the range modulator.

Our approach consists of providing a map of the fluence $\bm{F}$ at the entrance of the range modulator to a Monte Carlo dose engine which was modified to perform and store a simulation independently for each pixel of the fluence map. Those developments are detailed in section~\ref{sec:mcsquare}. For each pixel $h_j$ of the range modulator, we compute the dose $D(\bm{w}, \bm{h})$ for protons originating only at this location and with initial particles at nominal energy $E$. The number of protons is proportional to $F_j$. We then compute the dose $D(\bm{w}, \bm{h})$ in the exact same conditions but an initial energy of $E - \Delta E$. The derivative of $D_i(\bm{w}, \bm{h})$ may then be approximated as:
\begin{eqnarray}
    \frac{\delta D_{ij}(\bm{w}, h_j)}{\delta E} &=& \frac{\delta D_i(\bm{w}, E_j)}{\delta E_j} \\
    &\approx& \frac{D_i(\bm{w}, E_j) - D_i(\bm{w}, E_j - \Delta E_j)}{\Delta E_j}\label{eq:finite_diff}
\end{eqnarray}

Despite the modifications made in the dose calculator, the execution time may remain significant. To mitigate this, we propose a preliminary estimation of the objective function gradient using the continuous slowing down approximation (CSDA). Specifically, we estimate the dose in each voxel based on the energy loss calculated from a 3D map of relative proton stopping powers (RSPs) as follows:

\begin{eqnarray}
D_{x,y, z} &=& F_{x, y} SP_{w} \left( R(E_0) - \int RSP_{x, y, z} dz \right)
\end{eqnarray}

Here, $z$ is assumed to be along the beam direction, $SP_{w} (R(E))$ is the mass stopping power of water with respect to the range of protons having an energy equal to $E$, and $E_0$ is the nominal energy of the beam. While this approach neglects important phenomena like scattering and beam divergence, it can efficiently provide a rough estimate of $\mathbf{h}$. It is worth noting that this simple analytical approach often yields satisfactory results for the values of $\mathbf{h}$.

\subsection{Monte Carlo computations based on fluence map\label{sec:mcsquare}}
The Monte Carlo dose engine used in this study is MCsquare\cite{Souris2016} which takes advantage of mutli-core CPU architectures. MCsquare has the ability to compute the dose independently for each beamlet and store it in a sparse matrix without significant increase of the computation time. We hacked this feature to store the dose associated to each pixel of the range modulator.

MCsquare takes as input the location of the spot on the isocenter plane. To provide the fluence map to MCsquare, we first estimated the fluence at the entrance of the range modulator, given a model of the beam. Then, we projected the coordinates of the fluence values onto the isocenter plane given the distance between the steering magnets and the isocenter. Finally, the fluence map at isocenter was simply converted to the classical input taken by MCsquare.

Some modifications were made to the code of MCsquare to bypass its default sampling scheme. By default, MCsquare samples the initial particle locations according to a probability density function of which the parameters are defined in a beam model given as input. This default behavior was changed so that the initial location would be exactly the position of the pixels of the fluence map projected onto the plane of the nozzle exit.

In our modified MCsquare, the amount of particles to sample at each $(x, y)$ is now considered a deterministic variable corresponding to the pixels of the fluence map. The correlation between the particle direction and its location with respect to the center of the spot was also removed as such spots are not used anymore. Mathematically, MCsquare uses the following covariance matrix $\Sigma^2$ to sample the particles at the nozzle exit, for each beamlet:
\begin{equation}
\Sigma (z_{nozzle}) = 
  \begin{pmatrix}
    \sigma_x(z_{nozzle}) & \rho_{x \theta}(z_{nozzle})\\
    \rho_{x \theta}(z_{nozzle}) & \sigma_{\theta}(z_{nozzle})
  \end{pmatrix}  
\end{equation}
where $\sigma_x(z_{nozzle})$ is the standard deviation for the particle location, $\sigma_{\theta}(z_{nozzle})$ is the standard deviation for the particle direction and $\rho_{x \theta}$ is the correlation between particle position and direction. A similar covariance matrix is used for the y axis, de facto assuming no correlation between x and y.

However, when we provide the fluence at the nozzle exit, the initial location of the particle is now deterministic. It is constant for each pixel of the fluence map and therefore $\sigma_x(z_{nozzle})$ is zero. Consequently, it does not make sense anymore to consider any correlation between the particle direction and its constant initial location. The covariance matrix implemented in MCsquare was thus changed to:
\begin{equation}
\Sigma (z_{nozzle}) = 
  \begin{pmatrix}
    0 & 0\\
    0 & \sigma_{\theta}(z_{nozzle})
  \end{pmatrix}  
\end{equation}

\subsection{Optimization of the PBS spot weights}
The optimization of the spot weights is done classically as in Barragan~et~al.\cite{Barragan2018}. A dose influence matrix is first computed using MCsquare. The optimization problem~\ref{eq:opti_problem} is then solved by L-BFGS.

\subsection{Range modulator representation and insertion within the CT}
The range modulator is considered as a pixelated object which can be placed directly in the CT. The thickness in $(x, y)$ of the range modulator is encoded in a pixel value, as shown on Fig.~\ref{fig:schem_CEF}. In 3D, the range modulator can be seen as a collection of 'towers' attached to a plain area which can be considered as a range shifter. Those two kinds of components can be made in different materials. Although the use of plastic material for the range modulator fabrication would be convenient, a higher density material like aluminum could be chosen for the range shifter instead. This would serve to improve the lateral dose uniformity by increasing the amount of lateral scattering.

Splitting the range modulator into 'towers' and a range shifter is done dynamically at each iteration of the optimization. A minimum thickness of plain material is left attached to the 'towers' accordingly to 3D printing specifications which might be provided by the manufacturer. In this study this thickness was set to 5~mm.

To avoid resampling artefacts of the range modulators, all computations were done on CTs resampled on the beams-eye views. The CT resolution was $1\times1\times2~\mathrm{mm^3}$. Calculated dose maps were then resampled back on the original CT.

\subsection{Experimental validation}
To evaluate our optimization method, we used it to design conformal FLASH plans on an head and neck case. These plans were compared to conventional IMPT. Three PTVs were drawn as extentions of the CTVs: one on the left side and two on the right side of the patient. The prescriptions were 54.25~Gy for the left PTV and 54.25~Gy and 70~Gy for the two PTVs on the right side of the patient.

IMPT plans were optimized using RayStation~11b (RaySearch Laboratories, Stockholm, Sweden) considering four fields with gantry angles of around 60\degree, 120\degree, 240\degree, and 300\degree, and 10\degree couch rotation. A range shifter of 40 mm equivalent to water was used for all fields.

As FLASH treatments will most likely be hypofractionated\cite{Van_de_Water2019-pk} and a long delay might be required between the delivery of each field, setup errors might be critical. Designing the range modulator within a robust optimization framework would be highly desirable. However, this must be considered in a second instance after that the dose conformity which can be achieved with a range modulator is proved to be sufficient. Consequently, in the absence of a robust framework, the optimization of each field was done so as to have a uniform dose for that field. Given the shape of the target volume, it is clear that 120\degree and 240\degree are not appropriate to have a uniform dose. Therefore, those beam angles were not considered for FLASH treatment optimization which suggests that a different approach will have to be taken when planning FLASH treatments or to select patients eligible for a FLASH treatment.

A Python implementation of the aforementioned method was utilized to optimize FLASH plans. The maximum energy attainable from an IBA system, 226 MeV, was selected as the energy for the FLASH treatment. A spot size, characterized by a Gaussian distribution with standard deviations of $\sigma_x = 4.5$ mm and $\sigma_y = 5$ mm, was chosen. To account for the amplified scattering in the FLASH configuration, a spot spacing of 7~mm was employed in the present investigation.

The materials used for the range modulator, the range shifter and the collimator were respectively: PMMA (density: $1.2~\mathrm{g/cm^3}$), aluminum (density: $2.7~\mathrm{g/cm^3}$) and tungsten (density: $19.3~\mathrm{g/cm^3}$).

\section{Results\label{sec:results}}
\subsection{Dose conformity}
\begin{figure}
    \centering
    \includegraphics[width=0.49\textwidth]{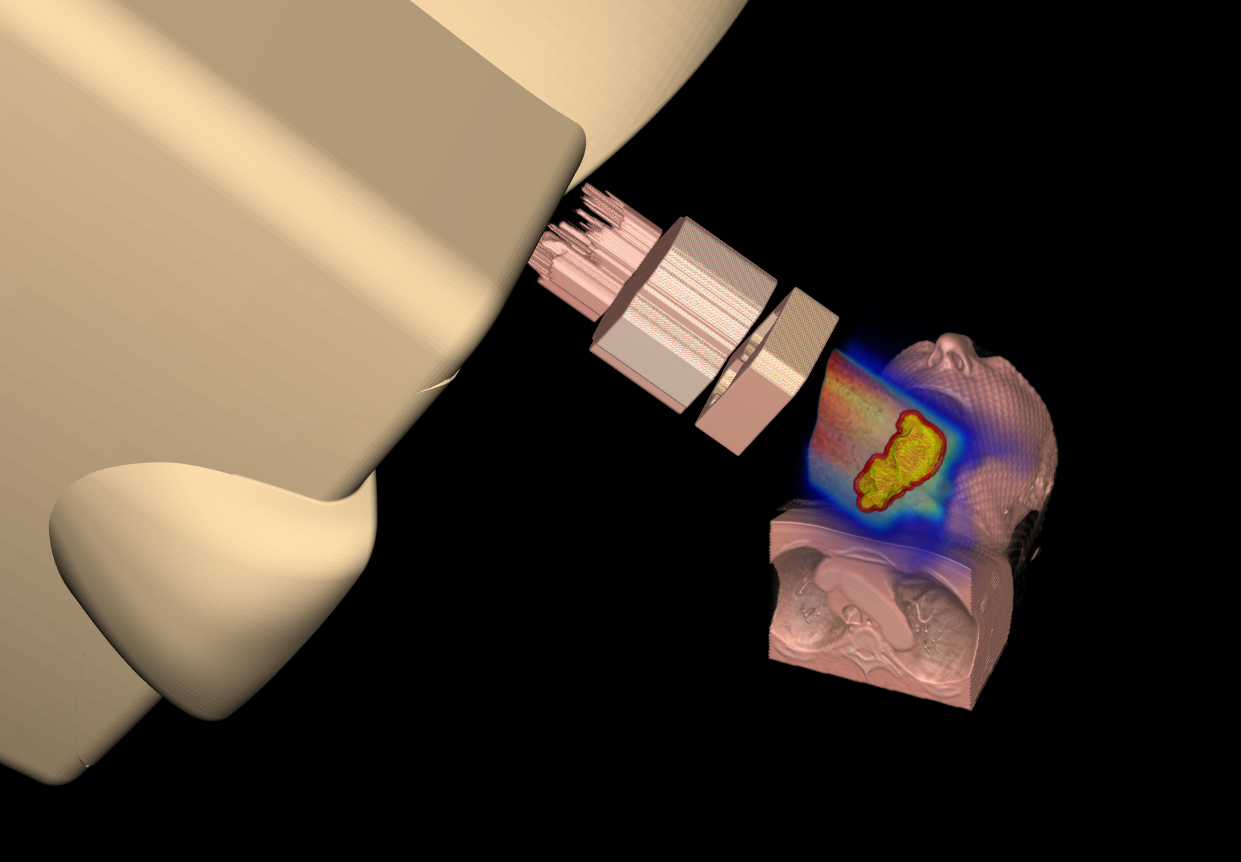}
    \caption{Experimental setup for the right part of the tumor. From nozzle to patient: range modulator (PMMA), range shifter (aluminium), and aperture (tungsten).}
    \label{fig:ptv7000_setup}
\end{figure}

To evaluate our optimization method, we designed a conformal FLASH plan on an head and neck case. The target was composed of two rather large volumes, with irregular contours and variable depths. Fig.~\ref{fig:ptv7000_setup} shows the experimental setup and the dose distribution for the right part of the tumor.

In this study, we mainly focused on the capabilities of the algorithm to give a uniform dose inside the target. We could not carry out an exhaustive dosimetric study as this head and neck case is not suitable for treatment with two opposite fields each giving a uniform dose. The combination of these fields would inevitably contribute to a superimposition of dose around the pharynx and therefore to a hot spot. In addition, a comprehensive dosimetric study should not only address differences in dose distributions but also take into account the benefits of using a high dose rate.

\begin{figure*}
    \centering
    \includegraphics[width=\textwidth]{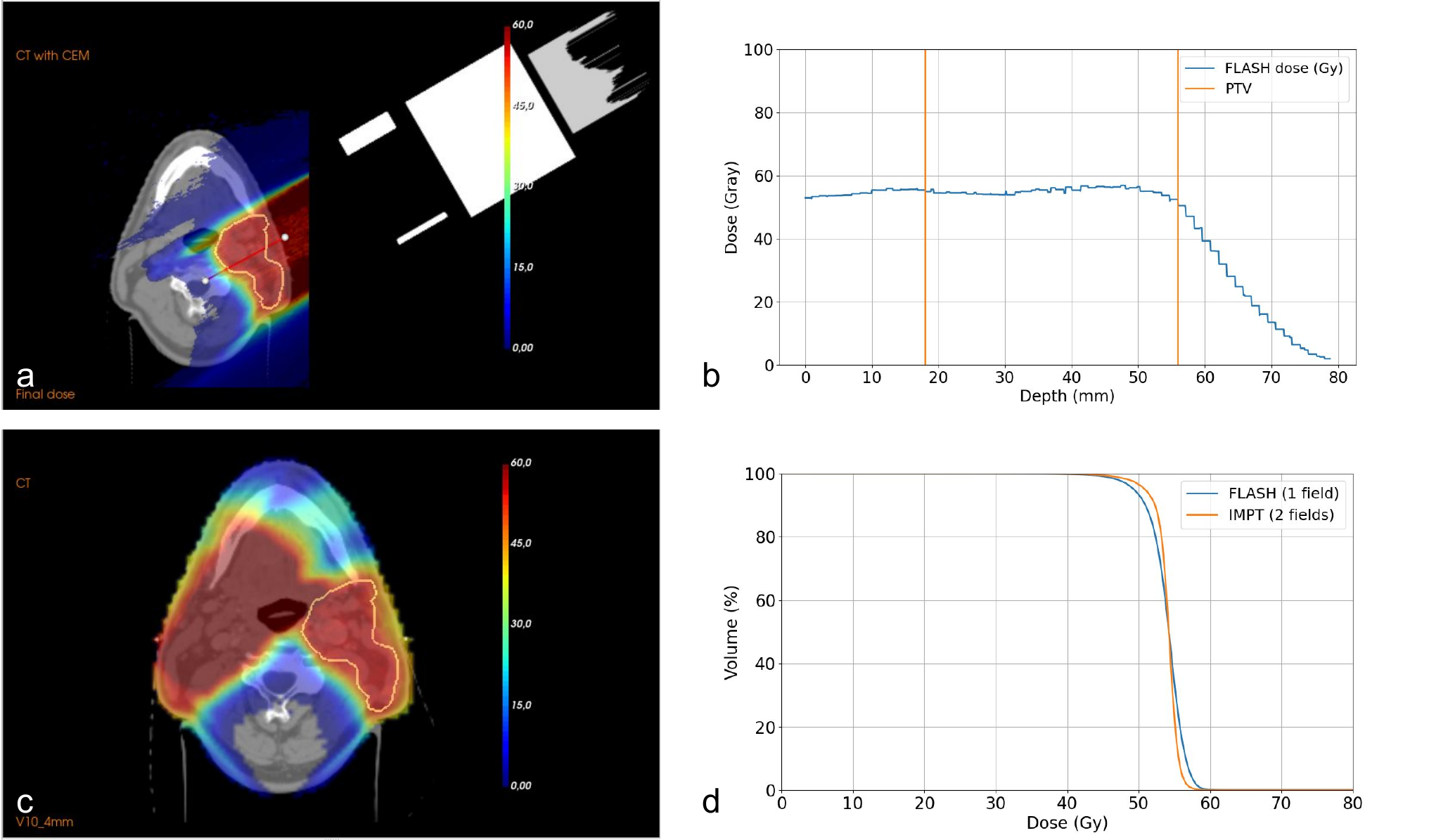}
    \caption{(a) Dose distribution obtained with the FLASH plan, (b) dose profile corresponding to a line passing through the center of the PTV, (c) IMPT dose distribution, and (d) DVH comparison for the left PTV and a prescription of 54.25~Gy.}
    \label{fig:imptComparison5425}
\end{figure*}

\begin{figure*}
    \centering
    \includegraphics[width=\textwidth]{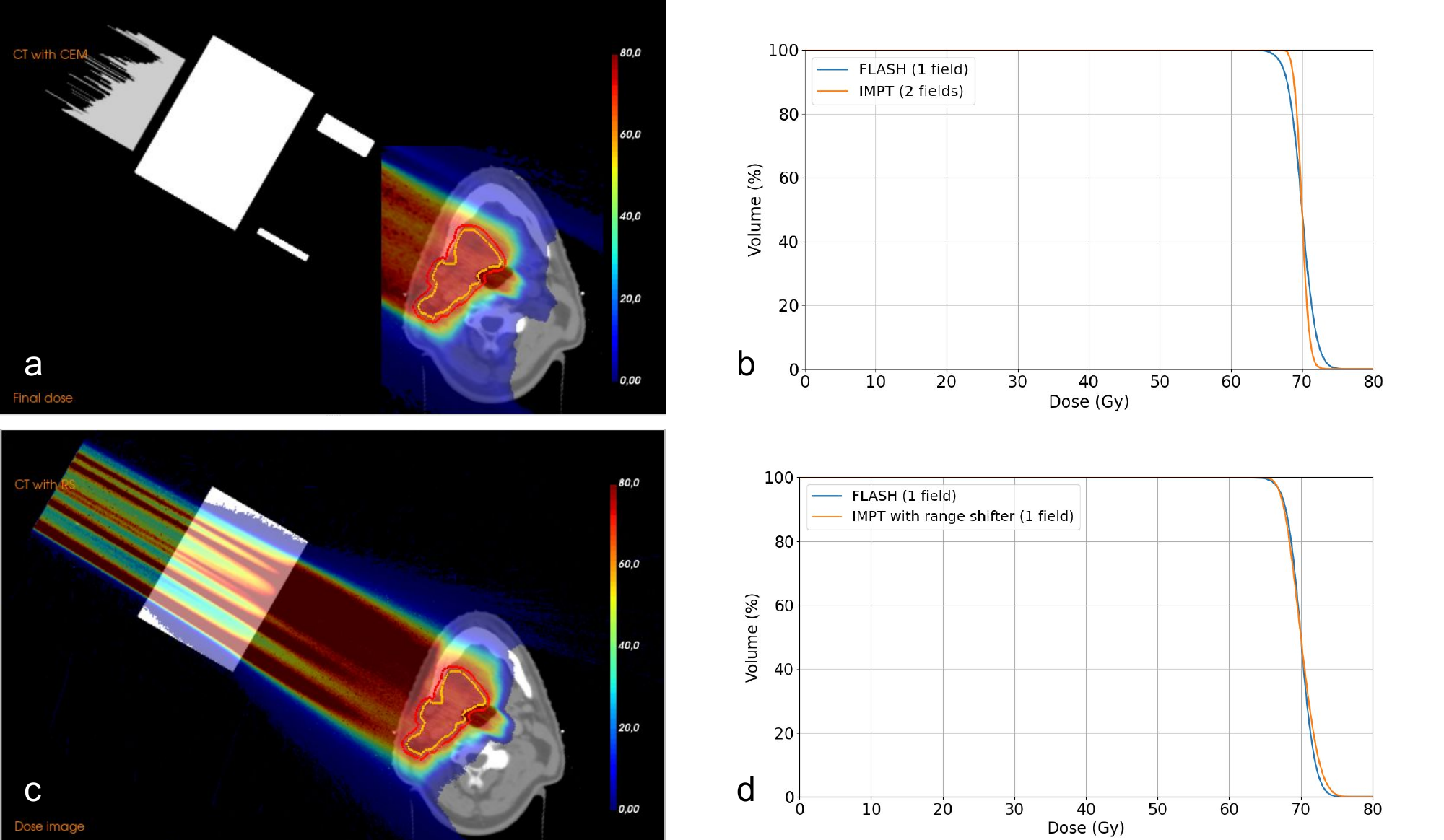}
    \caption{(a) Dose distribution obtained with the FLASH plan, (b) comparison of the corresponding DVH and that obtained in standard IMPT, (c) Dose distribution obtained with the single field IMPT plan and the same range shifter as for the FLASH plan, and (d) comparison of the corresponding DVH and that obtained in FLASH, in the CTV.}
    \label{fig:imptComparison7000}
\end{figure*}

A Comparison with IMPT is shown on Fig.~\ref{fig:imptComparison5425} and~\ref{fig:imptComparison7000}. For the left PTV, the dose uniformity was close to IMPT. In the PTV we had $D95 = 52.5~\mathrm{Gy}$ and $D5 = 57.9~\mathrm{Gy}$ in FLASH, and $D95 = 51.8~\mathrm{Gy}$ and $D5 = 57.1~\mathrm{Gy}$ in IMPT. The dose was less uniform in the right PTV with $D95 = 65.2~\mathrm{Gy}$ and $D5 = 74.5~\mathrm{Gy}$ in FLASH, and $D95 = 67.3~\mathrm{Gy}$ and $D5 = 71.4~\mathrm{Gy}$ in IMPT. However, IMPT dose maps were obtained with two fields for each PTV whereas the FLASH treatment only had one field for each PTV.

To determine whether the degradation in dose uniformity was caused by the use of the range modulator or rather by the use of a single field and a large range shifter, we removed the range modulator from the CT and designed an IMPT plan with a single field. Fig.~\ref{fig:imptComparison7000}c shows the experimental setup: the range modulator was removed but the range shifter was left at the exact same place in the CT. The spot spacing used for that IMPT plan was the same as that used to design the FLASH plan. In Fig.~\ref{fig:imptComparison7000}d, the comparison of the dose profiles and the dose-volume histograms (DVHs) in the CTV shows that the design of the range modulator was actually very close to being optimal for these conditions, and that the dose degradation mainly came from the use of a single field and a large range modulator. D95s differ by $0.1~\mathrm{Gy}$ and D5s differ by $0.7~\mathrm{Gy}$.

\subsection{Dose rate}
\begin{figure*}
    \centering
    \includegraphics[width=\textwidth]{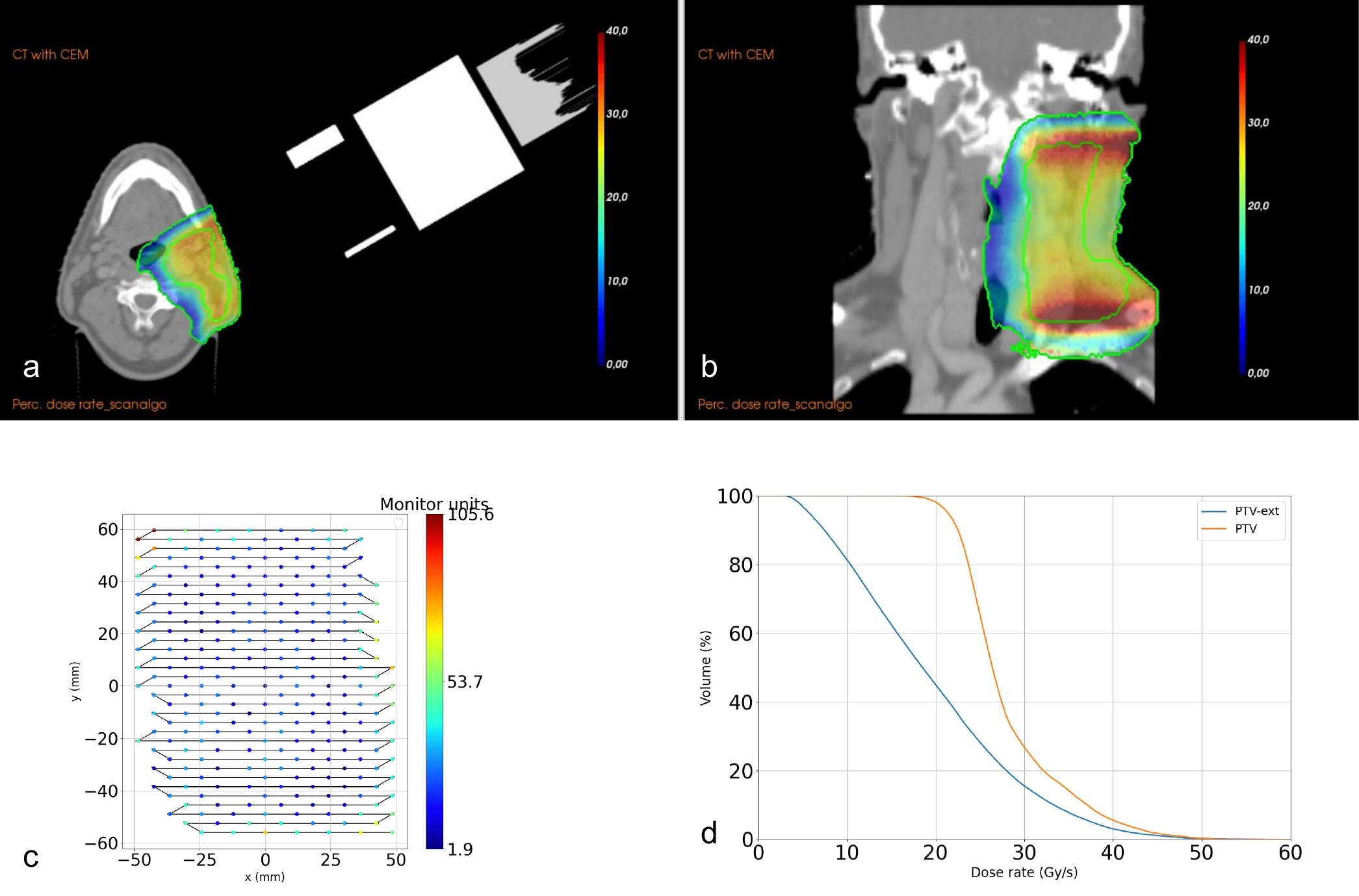}
    \caption{(a) and (b) Dose rate distribution computed for a current of 320 nA, (c) map of the intensity of the PBS spots, and (d) dose rate volume histogram.}
    \label{fig:drvh}
\end{figure*}

In the current study, the PTVs considered had prescription doses of 54.5 Gy and 70 Gy. However, to accurately compute the dose rate, it is imperative to consider the dose per fraction, which is likely to be around 8~Gy for hypofractionated FLASH treatments. The spots were arranged in a conventional serpentine pattern, as depicted in Fig.~\ref{fig:drvh}c. The scanning speed was set at 8000 mm/s, and the current averaged on a pulse period was 500 nA at the nozzle output.

To assess the efficacy of a FLASH treatment, it is pertinent to evaluate the dose rate in the organs at risk (OARs) and the healthy tissues surrounding the target volume. However, to facilitate a direct comparison, a unique volume surrounding the PTV was defined instead of distinct volumes for each OAR. To delineate this volume, we first computed an extension of the PTV by 20 mm, named PTV-ext, which excluded areas where the dose was lower than 1 Gy.

The dose rate maps, along with the dose rate volume histograms (DRVHs), are presented in Fig.~\ref{fig:drvh}. It can be observed that the threshold of 40 Gy/s, commonly associated with the FLASH effect\cite{Favaudon2014}, is not achieved. Notably, the PBS dose rate is influenced by the length to be scanned in the primary scanning direction. Thus, the dose rate may potentially be enhanced through optimization of the scanning pattern.

\section{Discussion}
The results presented in the previous section showed that the range modulators which were optimized for a complex head and neck case were optimal when considering dose uniformity and conformality.

One of the novel aspects in the proposed approach is that the range modulator is voxelized which has several advantages. First, it can be placed in the planning CT for dose computation. This greatly simplifies the simulation workflow as an external Monte Carlo dose engine supporting parameterized geometries like Geant4 do not need to be used. In addition, the geometry of the range modulator is arbitrarily set by the optimizer to optimally conform to the complex shape of the target volume. The algorithm is free to determine the height of each individual pixel of the elevation map and this typically results in asymmetrical pyramids as can be seen on the figures presented in Section~\ref{sec:results}.

Arbitrary geometry was not the sole improvement provided by the proposed method. Both the weights of the PBS spots and the individual heights of the range modulator were optimized directly under constraints. Thus, the range modulator was not optimized given spot weights but instead both spot weights and modulator elevations were updated at each iteration as dependent variables. This was possible thanks to the use of a fast Monte Carlo which could discriminate the contribution of each pixel of the range modulator.

In addition to the advantages just mentioned above, the proposed voxelized range modulator has a rather coarse resolution compared to other works in which it is similar to a conventional ridge filter with a high number of pyramidal peaks of variable prominence. The proposed range modulator typically has a resolution of around $1\times1\times1~\mathrm{mm^3}$. This is an asset for 3D printing. Even though current printing techniques have a precision of much less than one millimeter, this is not necessarily desirable from a mechanical point of view, in particular regarding the geometrical integrity in the presence of vibration or any externally applied force, including gravity.

In this first study, we compared a conformal FLASH plan to an IMPT plan optimized with a modern TPS considering robustness criteria. Conformal FLASH logically underperformed when compared to IMPT which comes from physical limitations, in particular the degradation of the distal fall off. However, we showed that the presented approach led to a dose distribution very close to that obtained with a single field IMPT and the same range shifter as that used for FLASH. The remaining slight difference could be explained by the fact that just as the range shifter degrades the sharpness of the Bragg peak, an additional degradation arises from the use of the range modulator. These observations makes it clear that the single field uniform dose approach considered in FLASH cannot compete with conventional IMPT in terms of dose conformity, and that should drive the selection of patient for FLASH treatments and the way that the plan would be designed. In addition, the optimization method could be integrated into a more complete approach, in particular with robustness criteria and a model of the FLASH effect on tumor control.

\section{Conclusions}
A joint range modulator and treatment plan optimization method was proposed and validated on an head an neck case. The optimization is done directly under constraints, similarly to IMPT inverse planning. The range modulator has a voxilized geometry which has many advantages: a simpler dose calculation pipeline, a geometry which can be arbitrarily determined by the optimizer and which is well suited for 3D printing.

\section{Acknowledgments}
This work was supported by the Walloon Region of Belgium through technology innovation partnership no. 8341 (EPT-1 – Emerging Proton Therapies Phase 1) co-led by MecaTech and BioWin clusters.

\section{Conflict of Interest Statement}
Dr. Kevin Souris was an employee of Ion Beam Applications during the writing of this paper.

\bibliographystyle{plain}

\begin{thebibliography}{10}

\bibitem{Barragan2018}
Ana~María Barragán~Montero, Kevin Souris, Daniel Sanchez-Parcerisa, Edmond
  Sterpin, and John~Aldo Lee.
\newblock Performance of a hybrid monte carlo-pencil beam dose algorithm for
  proton therapy inverse planning.
\newblock {\em Medical Physics}, 45(2):846--862, 2018.

\bibitem{BEYREUTHER2019}
Elke Beyreuther, Michael Brand, Stefan Hans, Katalin Hideghéty, Leonhard
  Karsch, Elisabeth Leßmann, Michael Schürer, Emília~Rita Szabó, and Jörg
  Pawelke.
\newblock Feasibility of proton flash effect tested by zebrafish embryo
  irradiation.
\newblock {\em Radiotherapy and Oncology}, 139:46--50, 2019.
\newblock FLASH radiotherapy International Workshop.

\bibitem{BUONANNO2019}
Manuela Buonanno, Veljko Grilj, and David~J. Brenner.
\newblock Biological effects in normal cells exposed to flash dose rate
  protons.
\newblock {\em Radiotherapy and Oncology}, 139:51--55, 2019.
\newblock FLASH radiotherapy International Workshop.

\bibitem{Favaudon2014}
Vincent Favaudon, Laura Caplier, Virginie Monceau, Frédéric Pouzoulet, Mano
  Sayarath, Charles Fouillade, Marie-France Poupon, Isabel Brito, Philippe
  Hupé, Jean Bourhis, Janet Hall, Jean-Jacques Fontaine, and Marie-Catherine
  Vozenin.
\newblock Ultrahigh dose-rate flash irradiation increases the differential
  response between normal and tumor tissue in mice.
\newblock {\em Science Translational Medicine}, 6(245):245ra93--245ra93, 2014.

\bibitem{JIN2020}
Jian-Yue Jin, Anxin Gu, Weili Wang, Nancy~L. Oleinick, Mitchell Machtay, and
  Feng-Ming {(Spring) Kong}.
\newblock Ultra-high dose rate effect on circulating immune cells: A potential
  mechanism for flash effect?
\newblock {\em Radiotherapy and Oncology}, 149:55--62, 2020.

\bibitem{LABARBE2020}
Rudi Labarbe, Lucian Hotoiu, Julie Barbier, and Vincent Favaudon.
\newblock A physicochemical model of reaction kinetics supports peroxyl radical
  recombination as the main determinant of the flash effect.
\newblock {\em Radiotherapy and Oncology}, 153:303--310, 2020.
\newblock Physics Special Issue: ESTRO Physics Research Workshops on Science in
  Development.

\bibitem{Liu2022}
Ruirui Liu, Serdar Charyyev, Niklas Wahl, Wei Liu, Minglei Kang, Jun Zhou,
  Xiaofeng Yang, Filipa Baltazar, Martina Palkowitsch, Kristin Higgins, William
  Dynan, Jeffrey Bradley, and Liyong Lin.
\newblock An integrated biological optimization framework for proton sbrt flash
  treatment planning allows dose, dose rate, and let optimization using
  patient-specific ridge filters, 2022.

\bibitem{MontayGruel2019}
Pierre Montay-Gruel, Munjal~M. Acharya, Kristoffer Petersson, Leila Alikhani,
  Chakradhar Yakkala, Barrett~D. Allen, Jonathan Ollivier, Benoit Petit,
  Patrik~Gonçalves Jorge, Amber~R. Syage, Thuan~A. Nguyen, Al~Anoud~D.
  Baddour, Celine Lu, Paramvir Singh, Raphael Moeckli, François Bochud,
  Jean-François Germond, Pascal Froidevaux, Claude Bailat, Jean Bourhis,
  Marie-Catherine Vozenin, and Charles~L. Limoli.
\newblock Long-term neurocognitive benefits of flash radiotherapy driven by
  reduced reactive oxygen species.
\newblock {\em Proceedings of the National Academy of Sciences},
  116(22):10943--10951, 2019.

\bibitem{MONTAYGRUEL2017}
Pierre Montay-Gruel, Kristoffer Petersson, Maud Jaccard, Gaël Boivin,
  Jean-François Germond, Benoit Petit, Raphaël Doenlen, Vincent Favaudon,
  François Bochud, Claude Bailat, Jean Bourhis, and Marie-Catherine Vozenin.
\newblock Irradiation in a flash: Unique sparing of memory in mice after whole
  brain irradiation with dose rates above 100gy/s.
\newblock {\em Radiotherapy and Oncology}, 124(3):365--369, 2017.
\newblock 15th International Wolfsberg Meeting 2017.

\bibitem{Oelfke2001}
U~Oelfke and T~Bortfeld.
\newblock Inverse planning for photon and proton beams.
\newblock {\em Med Dosim}, 26(2):113--124, 2001.

\bibitem{Pratx2019}
Guillem Pratx and Daniel~S Kapp.
\newblock A computational model of radiolytic oxygen depletion during {FLASH}
  irradiation and its effect on the oxygen enhancement ratio.
\newblock {\em Physics in Medicine \& Biology}, 64(18):185005, sep 2019.

\bibitem{Rothwell2022}
Bethany Rothwell, Matthew Lowe, Erik Traneus, Miriam Krieger, and Jan
  Schuemann.
\newblock Treatment planning considerations for the development of flash proton
  therapy.
\newblock {\em Radiotherapy and Oncology}, 2022/08/25 XXXX.

\bibitem{Simeonov2022}
Yuri Simeonov, Uli Weber, Christoph Schuy, Rita Engenhart-Cabillic, Petar
  Penchev, Veronika Flatten, and Klemens Zink.
\newblock Development, monte carlo simulations and experimental evaluation of a
  3d range-modulator for a complex target in scanned proton therapy.
\newblock {\em Biomedical Physics \& Engineering Express}, 8(3):035006, mar
  2022.

\bibitem{Souris2016}
Kevin Souris, J~Lee, and E~Sterpin.
\newblock Fast multipurpose monte carlo simulation for proton therapy using
  multi‐ and many‐core cpu architectures.
\newblock {\em Medical Physics}, 43(4):1700--1712.

\bibitem{Spitz2019}
Douglas~R. Spitz, Garry~R. Buettner, Michael~S. Petronek, Jo{\"e}l~J. St-Aubin,
  Ryan~T. Flynn, Timothy~J. Waldron, and Charles~L. Limoli.
\newblock An integrated physico-chemical approach for explaining the
  differential impact of flash versus conventional dose rate irradiation on
  cancer and normal tissue responses.
\newblock {\em Radiotherapy and Oncology}, 139:23--27, Oct 2019.

\bibitem{Van_de_Water2019-pk}
Steven van~de Water, Sairos Safai, Jacobus~M Schippers, Damien~C Weber, and
  Antony~J Lomax.
\newblock Towards {FLASH} proton therapy: the impact of treatment planning and
  machine characteristics on achievable dose rates.
\newblock {\em Acta Oncol}, 58(10):1463--1469, June 2019.

\bibitem{Venkatesulu2019}
Bhanu~Prasad Venkatesulu, Amrish Sharma, Julianne~M Pollard-Larkin, Ramaswamy
  Sadagopan, Jessica Symons, Shinya Neri, Pankaj~K Singh, Ramesh Tailor,
  Steven~H Lin, and Sunil Krishnan.
\newblock Ultra high dose rate (35 gy/sec) radiation does not spare the normal
  tissue in cardiac and splenic models of lymphopenia and gastrointestinal
  syndrome.
\newblock {\em Sci Rep}, 9(1):17180, November 2019.

\bibitem{Zhang2022}
Guoliang Zhang, Wenchao Gao, and Hao Peng.
\newblock Design of static and dynamic ridge filters for flash–impt: A
  simulation study.
\newblock {\em Medical Physics}, 49(8):5387--5399, 2022.

\end{thebibliography}

\end{document}